\documentclass[%
 amsmath,amssymb,
 aps,
 prl,
 twocolumn,
]{revtex4-1}
\usepackage{dcolumn,color,helvet,times}
\usepackage{amsmath,lastpage,bm,textcomp}

\usepackage{graphicx}
\usepackage{epstopdf}
\usepackage{bm}        
\usepackage{amssymb}   
\usepackage{chemarr}
\begin{document}

\setcounter{page}{1} 
\title{A theoretical model for attachment lifetimes of kinetochore-microtubules: {Mechano-kinetic ``catch-bond'' mechanism for error-correction.}}

\author{Blerta Shtylla}
 \altaffiliation[Current Address:]{ Department of Mathematics and Statistics, Mount Holyoke College, South Hadley, MA} 
  \email{bshtylla@mtholyoke.edu}
\author{Debashish Chowdhury}%

\affiliation{ Mathematical Biosciences Institute, The Ohio State University, 1735 Neil Avenue, Columbus, OH 43210, U.S.A.\\
Department of Physics, India Institute of Technology, Kanpur India
}%

\pagestyle{headings}


\begin{abstract}
Before cell division, two identical copies of chromosomes are pulled apart by microtubule (MT) filaments that approach the chromosomes from the opposite poles a mitotic spindle. Connection between the MTs and the chromosomes are mediated by a molecular complex called kinetochore.  An externally applied tension can lead to detachment of the MTs from the kinetochore; the mean lifetime of such an attachment is essentially a mean first-passage time. In their {\it in-vitro} pioneering single-kinetochore experiments, Akiyoshi et al. (Nature {\bf 468}, 576 (2010)), observed that the mean lifetimes of reconstituted MT-kinetochore attachments vary non-monotonically with increasing tension. The counter-intuitive stabilization of the attachments by small load forces was interpreted in terms of a catch-bond-like mechanism based on a phenomenological 2-state kinetic model. Here we develop the first detailed microscopic model for studying the dependence of the lifetime of the MT-kinetochore attachment on (a) the structure, (b) energetics, and (c) kinetics of the coupling. The catch-bond-like mechanism emerges naturally from this model. Moreover,  {\it in-silico} experiments on this model reveal further interesting phenomena, arising from the subtle effects of competing sub-processes, which  are likely to motivate new experiments in this emerging area of single-particle biophysics.
\end{abstract}
\maketitle

Chromosome segregation by the mitotic spindle is one of the most important intracellular processes in eukaryotic cells 
\cite{karsenti01,smith04,bouck08,mogilner10}. Connections between chromosomes and microtubules (MT) are mediated by kinetochores, which are complex macromolecular structures \cite{santaguida09,mcintosh02,smith05b,welburn08,nogales11}.  
Due to the difficulties of isolating kinetochores from cells, the identification and spatial organization of the molecular components of kinetochores has posed significant challenges.
Recent high resolution imaging has provided an indication of the distribution of these components and even their stoichiometries \cite{joglekar09,wan09,lawrimore11}. 
Kinetochores form dynamic, and yet sufficiently strong, coupling with MTs that undergo stochastic transitions between growth and shortening. This interaction between kinetochore elements and the attached kinetochore MT (kMT) generates movement of the chromosome.
While a possible architecture of this nano-device is beginning to emerge, the mechanism by which it couples forces from kMT polymerization/depolymerization with chromosome movement is a major unresolved question with significant implications \cite{joglekar10,bloom10a,asbury11}.
A fundamental biophysical question in this context is: how the dynamics of the kMTs and externally applied tension (load force) affect the stability of the kMT-kinetochore coupling. 

Recent {\it in-vitro} experiments with reconstituted kinetochore-MT attachments in budding yeast \cite{akiyoshi10, gonen12} have provided evidence that MT kinetics and load forces can combine in unexpected ways. Strikingly, it was found that a limited range of forces can be more favorable for maintaing kinetochore attachment, whereby load selectively stabilizes attachment \cite{akiyoshi10}.

In this letter, we develop a detailed microscopic model that, to our knowledge, is the first theoretical analysis of this phenomenon, at the single kinetochore level. The mean lifetime of the MT-kinetochore attachment is essentially a mean first-passage time \cite{redner}. Calculating this mean first-passage time using our microscopic model, we investigate the dependence of the mean attachment lifetime on (i) the structure, (ii) energetics, and (iii) kinetics of the MT-kinetochore coupler. Akiyoshi et al.~\cite{akiyoshi10} argued that their counter-intuitive data are ``reminiscent of `catch-bonds''' that can be explained in terms of a phenomenological two-state kinetic model. A catch-bond-like mechanism emerges naturally in the  theoretical framework of our microscopic model as a consequence of the force-sensitivity of kMT depolymerization. Our results reveal wider varieties of trends of variation of the attachment time than those observed by Akiyoshi et al.~\cite{akiyoshi10}. We also indicate possible adaptations of the experimental techniques of Akiyoshi et al.\cite{akiyoshi10} that may be appropriate for testing our new predictions.

Almost all the theoretical models of MT-kinetochore coupling \cite{hill85b,joglekar02,molodtsov05,efremov07,mcintosh10,shtylla11}  are based exclusively on one of the two major mechanisms for force generation. In the biased-diffusion model, initially proposed by Hill \cite{hill85b}, the plus end of a kMT is assumed to be surrounded by a coaxial ``sleeve'' the inner surface of which is composed of several binding elements that bind specific kMT sites. The one-dimensional Brownian motion of the sleeve along the axis of the kMT is biased to increase overlap, because a larger number of kMT-sleeve bindings lowers the total energy of the system. The interplay of this biased diffusion and the depolymerization of the kMT gives rise to the  pull exerted by the coupler on the kinetochore. An alternative coupling mechanism is based on the ``power stroke'' exerted on a rigid ring by the curling protofilament tips of a depolymerizing MT \cite{molodtsov05,efremov07}.  
There is increasing recent structural evidence that kinetochores indeed engage kMTs through multivalent attachments that move along microtubules \cite{dong07,gonen12}. Therefore, a biased diffusion remains a valid candidate mechanism for MT-kinetochore coupling. But this evidences does not necessarily exclude a role of the well known curled tips of depolymerising MTs in the MT-kinetochore coupling.   

In contrast to most of the earlier theoretical work, the model we propose here is ``unified'' in the sense that it incorporates the key features of both these types of models. In our model the main elements of the biased diffusion model are treated explicitly. Moreover, the curling of the MT protofilament tips, a key feature of the power-stroke model, is captured implicitly by assuming a tension-induced slowing down of depolymerisation which is known to arise from the tension-induced suppression of the curling. Although force-mediated kMT alteration has been discussed in the literature \cite{nicklas88,skibbens93,skibbens95,rieder94,rieder98,inoue95,skibbens97,maddox03,cimini04}, modeling of these effects in the context of biased diffusion has been limited.

The theory we develop here is an extension of a one-dimensional force-based model of  a kMT-kinetochore ``coupler'' \cite{shtylla11}. In this extended version, the coupler is composed of multiple passive kinetochore elements that bind kMTs via a generalized biased-diffusion mechanism, along with active kinetochore force generators, all of which maintain dynamic attachment with shortening/growing attached microtubules. 

We begin with the simplified special version of our model that includes only the passive binders; later in this letter we incorporate also the active force generators. In the first simplified version,  the kinetochore coupler is modelled as a collection of binder element heads which represent the core binding area of a kinetochore. 

\begin{figure}[t]
\centering{
\includegraphics[width=3.2in,height=1.7in]{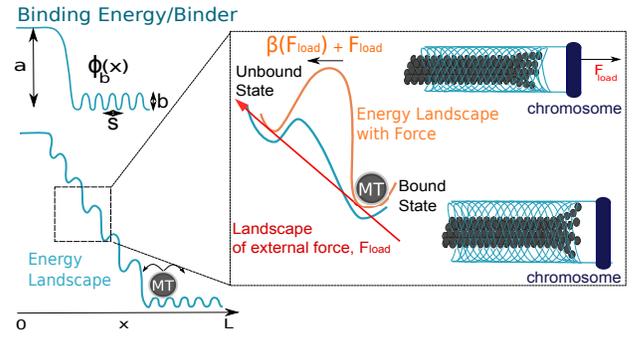}}
\caption{\small Diagram of model components.  $x=0$ is the kMT entry-point and $x=L$ is the maximum overlap. Mechano-kinetic modification leads to amplification of energy barriers for binding in a 'catch-bond' type mechanism.}
\label{fig1}
\end{figure}

The length of the overlap between the kMT and the coupler is denoted by $x$ (see Fig \ref{fig1}). Increasing the overlap between binders and the lattice is energetically favorable.  
As in previous work \cite{shtylla11}, we assume that each binder head engages with the kMT by obeying a unit energy function $\phi_{b}(x)$, which has two key parameters: $a$ measures free energy  drop due to binder affinity for the kMT lattice, and $b$ describes the activation barrier for transitions between specific kMT lattice binding sites. In other words, $a$ is a measure of the strength of the kMT-binder affinity while $b$ is a measure of the ``roughness'' of the kMT-coupler interface. The total potential energy function is given by 
\begin{equation}
\Psi_{b}(x)=\sum_{n}^{N_{b}}\phi_{b}(x-ns)
\end{equation}
where $s$ is the spacing between consecutive coupler binders (see Fig \ref{fig1}). Binder spacing is an arbitrary parameter (SI). Here we set $s=\ell$, where $\ell$ is the distance between consecutive kMT binding sites.

The coupler overlap velocity, $dx/dt$ is then given by the following stochastic differential equation
\begin{align}
dx(t)&=\frac{1}{\xi}\sum F dt \label{eq1} \\
&=\frac{1}{\xi}\left[ -\Psi_{b}'(x)-F_{\text{load}}\right]dt+ \ell dN_{r}(t) 
+\sqrt{2k_{B}T/\xi}dW(t)\nonumber,
\end{align}
where the constant $F_{\text{load}}$ is the external opposing load force on the coupler  and $\xi$ is the effective drag coefficient. $dW(t)$ accounts for the thermal diffusion of the coupler on the lattice, and $N(t)$ is poisson counting processes describing the kMT dynamics with intensity rate $r$.  

An opposing tension tends to decrease the overlap between the kMT and the coupler. We assume that the coupler under tension also suppresses the curvature of the tips of the kMT protofilaments so that the splaying tips of the kMT become confined within the coupler (we visualize the coupler protein meshwork as a children's finger trap toy, see Fig \ref{fig1}). Such force-induced suppression  of the curvature of the depolymerizing protofilament tips, in turn,  causes reduction of the rate of kMT depolymerization. This proposed scenario is consistent with the experimental observations of \cite{akiyoshi10}. Thus, kMT depolymerization rate is assumed to be a decreasing function of the load tension and the functional form of this dependence is assumed to be 
\begin{align}
 \beta(F)&=\beta_{\text{max}} e^{-\lambda F_{\text{load}}} 
 \end{align}
where the parameter $\lambda$ characterizes the extent of the effect of a given load tension on $\beta$.  The kMT polymerization rate $\alpha$, however, is assumed to be independent of load.  The rate functions are chosen in order to allow for the rate $r=\alpha-\beta(F)$ of the kMT tip dynamics to transition from a catastrophe to a rescue state in a load-dependent manner in agreement with observations in \cite{akiyoshi10} (SI). We define the breaking load $F_{\text{break}}$ to be the strength of the tension for which the mean attachment time is less than 1 min; the qualitative conclusions drawn on the basis of this definition do not depend sensitively on this choice.

The lifetime of a MT-kinetochore attachment is defined here mathematically as the time taken by the kMT, that is initially at $x=L$, to reach $x=0$ (the coupler entry point) {\it for the first time}. Since this time fluctuates from one MT-kinetochore attachment to another, we calculate the mean lifetime. The mean attachment lifetime is essentially a mean first passage time \cite{redner}, which we calculate using standard methods (SI). By a combination of analytical and numerical techniques, we study the trends of variation of the mean lifetime with (a) the strength of the externally applied load force, as well as, the (b) microscopic structure, (c) energetics, and (d) kinetics of the coupler.  $N_{b}$ is characteristic of the structure of the coupler (coupler length) whereas its energetics depend on $\Psi_{b}$ (i.e., on the parameters $a$, $b$) and $F_{\text{load}}$; the stochastic kinetics are influenced by the interplay of forces arising from the potential landscape, random Brownian forces, and by the kMT polymerization / depolymerization kinetics.



It is difficult to derive an exact analytical expression for the mean first passage time related to eq.~(\ref{eq1}). Therefore, we explore two limiting cases for which explicit approximate solutions can be obtained: (a) {\it  Slippery regime} (i.e., low-friction regime) where $b<<k_{B}T$; in this regime  the coupler can easily rearrange its position relative to the kMT, (b) {\it Strong friction regime}   where $b>>k_{B}T$; in this regime diffusion inside the binder is practically non-existent and  MT growth/ shortening rates are large compared to all other processes. Stronger friction weakens the ability of the coupler to quickly adjust its position with the variation of the length of a dynamic MT. For sufficiently large $b$, the coupler becomes static \cite{shtylla11}.

In the {\it slippery regime}, the mean lifetime is  (see SI for the derivation)
  \begin{align}
 T(L)\approx \frac{L^2}{D} \frac{\exp \left(-w\right)-1+w}{w^2} 
 \label{eq-diffTL}
 \end{align}
where $w=L(-a/\ell+F_{\text{load}}+\ell \beta_{max}\exp(-\lambda F_{\text{load}})\xi)/k_{B}T$ 
 is a dimensionless work quantity.  
 
\begin{figure}[t]
\begin{center}
\includegraphics[width=3.2in,height=2in]{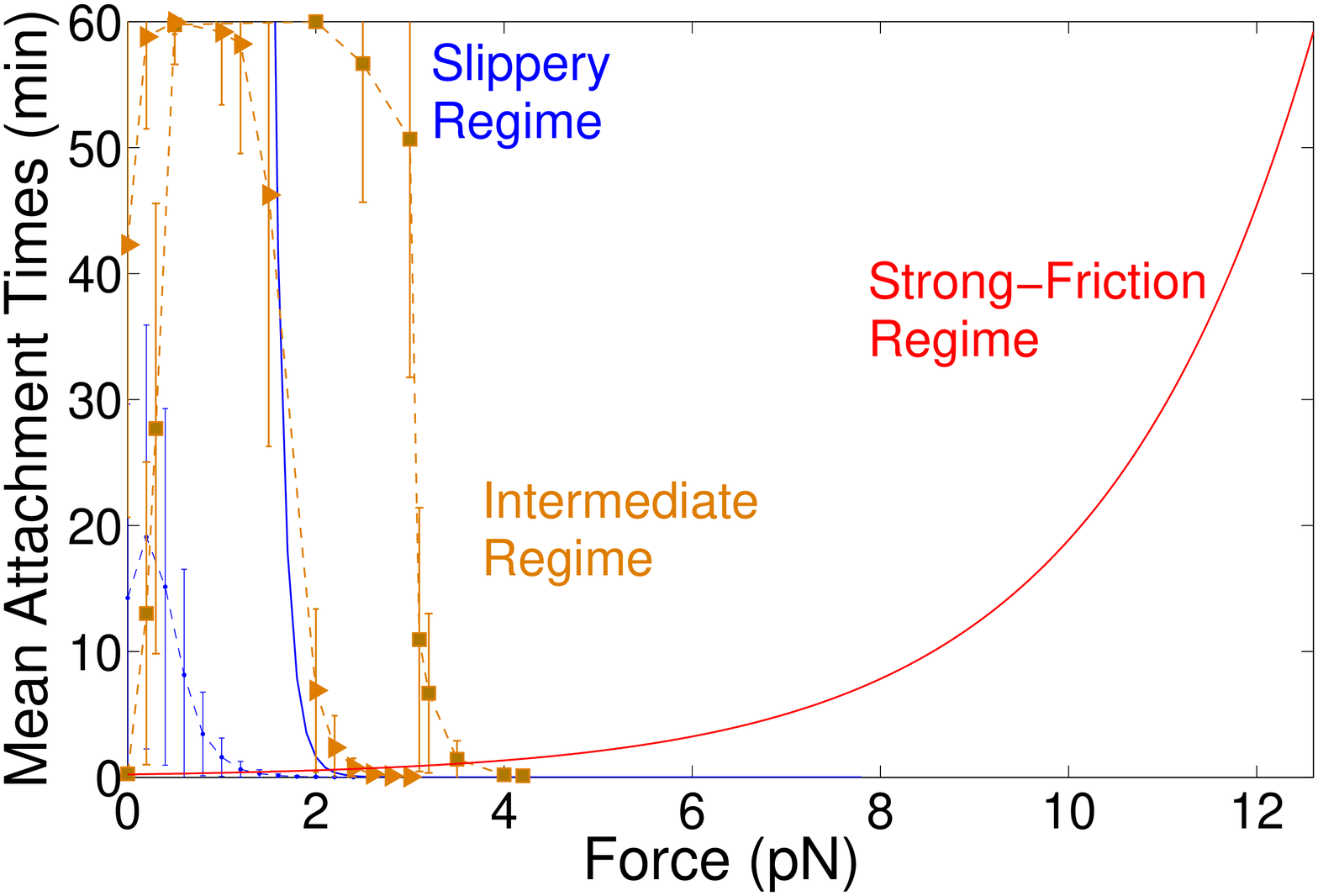}\hspace{-0.15in}
\caption{\small Mean attachment time versus load force. Blue, orange and red correspond to the slippery, intermediate and strong friction regimes, respectively. Solid blue line is obtained from eq.~(\ref{eq-diffTL})  with $\lambda=3 pN^{-1}, \alpha=40,\beta_{\text{max}}=120, N_{b}=65, a=0.4 k_{B}T, b=0.001a$; dashed blue line is obtained numerically for the same parameters but with $N_{b}=32$. 
Parameters for solid orange triangles  $\alpha=50,\beta_{\text{max}}=350, N_{b}=45, a=0.5 k_{B}T, b=0.04a$; those for solid orange squares are  same except that $a=0.6 k_{B}T, b=0.2a$. Solid red curve is obtained from eq. (\ref{eq-bindTL}) with $\beta_{\text{max}}=350, \alpha=0, \lambda=0.4$.} 
\label{fig2}
\end{center}
\end{figure}

 In the {\it strong friction regime}, the mean lifetime is (SI)
 \begin{equation}
  T(L)\approx
  \dfrac{L}{\ell\beta_{\text{max}} e^{-\lambda F_{\text{load}}}}. 
  \label{eq-bindTL}
  \end{equation}
In Fig \ref{fig2}, we show plots of mean lifetimes obtained by computer simulations for various parameter values, in addition to the expressions (\ref{eq-diffTL}) and (\ref{eq-bindTL}). Over a significant regime of physically relevant parameter values, our model gives rise to a non monotonic variation of the mean lifetime with the load tension; this is consistent with the experimental observation of Akiyoshi et al.\cite{akiyoshi10}. We have also explored other parameter regimes to understand kMT-kinetochore detachment phenomenon in further detail. The results indicate the possibility of other distinct trends of variation of the mean lifetime with tension that might be detectable in experiments under conditions different from those used by Akiyoshi et al.\cite{akiyoshi10}.

\begin{figure}[t]
\begin{center}
\includegraphics[width=3.4in,height=1.5in]{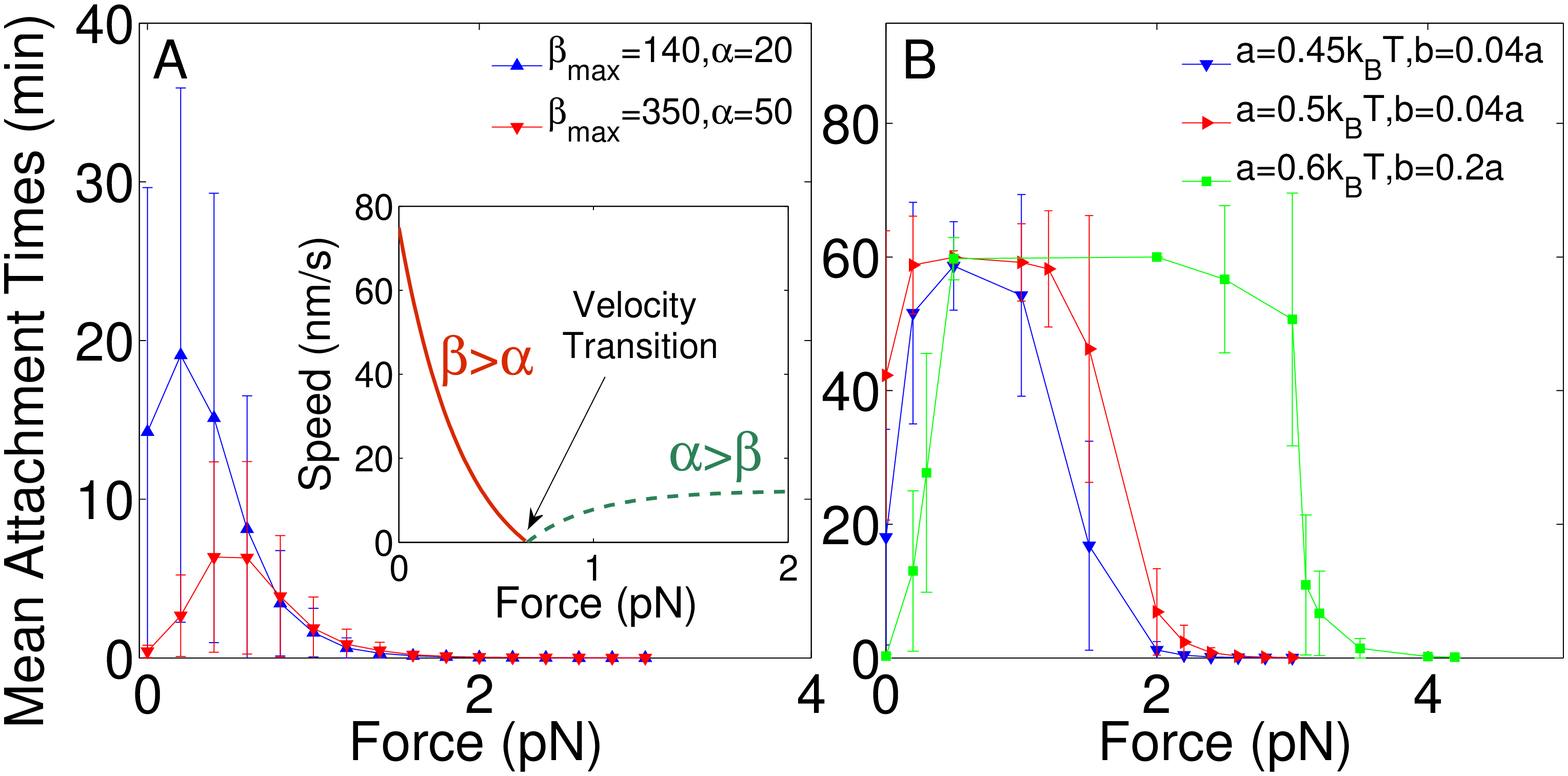}
\caption{\small  Mean attachment times for various coupler parameter regimes. {\bf A.} Mean attachment times in the {slippery regime} with different kMT kinetic rates for a short coupler with $N_{b}=32$. Common parameters are $\lambda=3$ pN$^{-1}$, $a=0.4$ k$_{B}$T, $k=0.001a$. 
 {\bf B.} Mean attachment time variations for an {\it intermediate} regime coupler with $N_{b}=45$ binders. An increase in the binding energies $a$ causes the tails of the mean attachment times to expand in larger force regions. 
 Common parameters are $\lambda=3$ pN$^{-1}$, $\beta_{max}=350$ s$^{-1}$, $\alpha=50$ s$^{-1}$. Error bars mark standard deviation.}
\label{fig3}
\end{center}
\end{figure}

\noindent {\it Results for the slippery regime}:
Figs.\ref{fig2} and \ref{fig3}A show that  in the {\it slippery regime} the trend of variation of the mean life time on the load tension depends sensitively on the binding energy (provided by $a$ or $N_{b}$). At sufficiently low binding energies, attachment times can be sensitive to depolymerization. Under these conditions, if depolymerization slows down by the load before the coupler breaks, then attachment times are non-monotonic and essentially follow $\beta(F)$ ($N_{b}=32$ in Fig.~\ref{fig2}).
The peak value of the mean attachment time depends on $\lambda=1/F_{c}$. Thus, for observing the non monotonic variation of the mean life time, $\lambda$ must be chosen so as to satisfy the requirement $F_{c}<F_{\text{break}}$.  Consequently, for  F- independent $\beta$ (which corresponds to the special case $\lambda=0$), the mean lifetimes decrease monotonically with increasing load tension, in agreement with \cite{akiyoshi10} (SI).  As our results establish, the range of breaking load $F_{\text{break}}$ can be easily adjusted for these model couplers by increasing the binding affinity for the lattice using $a$ or $N_{b}$ (two cases shown in Fig. \ref{fig2}). However, this enhanced stability comes at a price:  
in the stable slippery regime ($N_{b}=65$ in Fig.~\ref{fig2}), the lifetimes  for low loads increase beyond observable ranges and sharply decrease close to the breaking loads. 

\noindent {\it Results for the strong friction regime}: In this regime the diffusive motion of the coupler is made practically impossible by the condition $b \ll k_BT$. Moreover, increasing load tension cause stronger suppression of the kMT depolymerisation. Consequently, in this regime, the mean life time increases monotonically with increasing load tension (see Fig.\ref{fig2}).

\noindent {\it Results for the intermediate regime}:

We investigated the intermediate regimes numerically. The data in Fig. \ref{fig3} establish that, in this regime, molecular friction makes it harder for the kMT to exit the coupler under load, while it also further enhances the dependence of attachment time on kMT depolymerization.  A delicate balance between these effects strong friction and force leads to the observed trend of variation of mean life time with load force in this regime.  

The attachment regimes that result from our model reveal competing effects at the kinetochore sites. On one hand, load forces provide a pull that can detach the kMT from the kinetochore coupler. On the other hand, load force slows down depolymerization of the kMT, thereby slowing down the exit of its tip from the coupler which, effectively, counters the pulling effects of the load. In the conflict between these two opposing effects, as indicated by our data in the Figs. \ref{fig2} and \ref{fig3}, the internal coupler friction might be the ultimate determinant of the emergent behavior. In low and intermediate friction ranges, we distinguish a clear range of forces for which coupling is  selectively favored. This finding supports a force-mediated selective stabilization of kMT at the kinetochore coupler. We find that non monotonic variation of the mean life time with the tension, which resembles a a catch-bond mechanism similar to\cite{akiyoshi10} is just one of the possible responses of the kMT-kinetochore coupler. It might be possible to observe the  other types of theoretically predicted responses by creating the corresponding required conditions in the {\it in-vitro} experiments. Such conditions may be facilitated, for example, by biochemical modifications of the Ndc80 complexes at kinetochores \cite{cheeseman06,welburn10,umbreit12}, which are close candidates for our multivalent passive binders.

{\it Effects of force-generating motor proteins in the coupler}

In addition to the passive kMT binders, active force generating components also play an important role in maintaining and regulating kMT-kinetochore coupling. Among the active force generators, cytoskeletal motor proteins are believed to make the dominant contribution \cite{sharp00b,scholey03b,salmon04,yang07}. Using recent structural data \cite{dumont12,gonen12} we create a hybrid coupler, where the outermost layer is composed of passive components and the innermost layer closest to the chromosome is composed of an active interface (see Fig \ref{fig4}A).

To include these active components, we add an active force term in our model 
\begin{equation}
F_{\text{A}}(x)=d_{m}(x)(n_{-}f_{-}-n_{+}f_{+})
\end{equation}
where $d_{m}(x)$ measures the $x$-dependent  length of the active kinetochore interface. The parameters $n_{-}, n_+$ denote the average number of minus and plus end motors per unit length of MT embedded in the kinetochore structure while $f_{-}$ and $f_{+}$ denote the force generated by a single minus-end and plus-end directed motor, respectively. For each motor we postulate a linear force-velocity relation in agreement with previous work \cite{mogilner06, efremov07}.

A key feature of the hybrid coupler is that 
besides modifying the force balance of the coupler, active components also affect the internal friction coefficient  (see SI for details). \\
\begin{figure}[t]
\begin{center}
\includegraphics[width=3.in,height=2.3in]{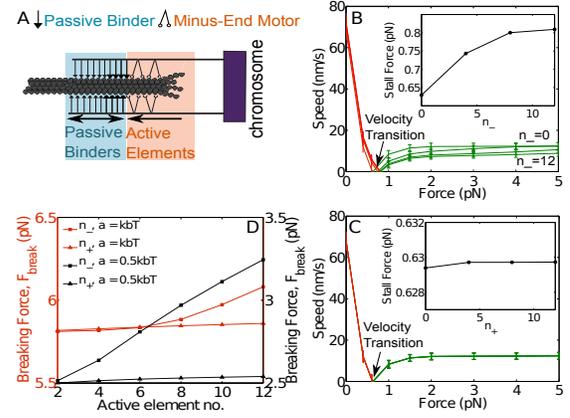}
\end{center}
\vspace{-0.2in}
\caption{\small 
{\bf A.} Diagram of the architecture of a hybrid coupler. {\bf B}. Numerical simulation results of average speed versus force for a stably bound coupler with varying densities of minus-end directed motors. {\it Inset}. Coupler stall forces for each motor density. {\bf C.} Average speed versus force for a stably bound coupler with varying densities of plus-end directed motors. {\bf D.} Breaking load calculations for the coupler shown in panel B-C for two binding energy values. The parameters used are $N_{b}=52$, $\lambda=3$, $\beta_{max}=130$ s$^{-1}$, $\alpha=20$ s$^{-1}$, $a=$ k$_{B}$T, $k=0.01a$. }
\label{fig4}
\end{figure}
In Fig. \ref{fig4} we plot force-velocity relations and $F_{\text{break}}$ for a hybrid coupler with varying numbers of motors at the active interface. The non-monotonic nature of mean attachment is preserved when the active components are added. When the motors oppose load force, stability of the kMT-kinetochore coupling is enhanced, with dynein (or protofilament curling)  being a strong candidate for this role. This effect is supported by the increased coupler breaking loads as $n_{-}$ increases, Fig. \ref{fig4}D. The active interface stabilizing effect of minus-end motors is particularly amplified when the binding energy of passive components is weakened, Fig \ref{fig4}D. Despite the stability of the attachment being improved, higher numbers of load opposing components also increase the internal friction of the coupler. This lowers the ability of the coupler to efficiently track rescued kMT tips, as noted by the slower coupler velocities in Fig. \ref{fig4}B.

Overlap-opposing motors (such as, CENP-E kinesin) under tension load bring the passive binders at the tip of the kMT.  We interpret these kinesins as components that increase the effective tension against the coupling which helps sustain tension-dependent suppression of depolymerization. Coupler velocities and $F_{\text{break}}$ do not show a significant difference as $n_{+}$ is varied under tension, Fig \ref{fig4}C. This suggests that these components do not significantly alter the nature of the kinetochore attachment if sufficient numbers of passive binders are engaged. Kinesins in the our model serve to enhance the kMT tip tracking efficiency of the coupler, rather than significantly destabilize passive coupling. Finally, when plus end motors experience compressing forces, motor friction effects dominate coupler movement (SI).


We conclude that the mean lifetime of the MT-kinetochore attachment depends on a very delicate balance of forces and kinetic effects. Over a range of parameter values, our model reproduces the non-monotonic variation of the lifetime with load force that was observed in \cite{akiyoshi10}. These attachment times give rise to optimal kMT-supported force ranges, indicating the existence of a force-dependent pathway for error correction at kinetochores. We emphasize that the catch-bond-like phenomenon, that arises naturally from our microscopic model, is only one of the several distinct responses of the kMT-kinetochore coupler to load tension. We have also explored other parameter  regimes that correspond to (a) different lengths of the coupler formed by the binders, (b) different potential landscapes, and (c) different rates of growth and shrinking of microtubules. We have also investigated the effects of a hybrid structure of the coupler that consists of an inner ``active'' layer of motor proteins and an outer ``passive'' layer of MT-binders. Depending on the parameter regime, the lifetime may appear to deviate from the qualitative features observed by Akiyoshi et al. \cite{akiyoshi10}. Some of the novel trends of variation observed in our analysis can be tested, in principle, by altering the size, composition, etc. of the single kinetochore particle in {\it in-vitro} experiments. \\
\href{https://www.mtholyoke.edu/~bshtylla/Articles/sc_short2_SI.pdf}{*Supplementary Information}
\vspace{-0.1in}
\section*{Acknowledgements}
{\small DC thanks MBI for hospitality. This research has been supported in part by the MBI, The Ohio State University and the National Science Foundation grant DMS 0931642 (BS and DC), DMS 1225251 (BS) and at IIT Kanpur (DC) by DBT (India), and by the Dr. J. M. Garg Chair professorship.} 


\pagebreak

\section{Supplementary Information}
\subsection{Kinetochore binder spacing: commensurate or incommensurate with MT lattice}

A key feature of our model is that the kinetochore fibrous corona is assumed to be densely populated by MT binder elements. Even though these elements can be flexible, for a dense enough kinetochore region, we can assume a regularly spaced array of binder elements that are connected to a rigid coupler backbone. This approach allows us to construct an energy landscape as a function of the number of binders, $N_{b}$ that can associate with the kMT. The binders are assumed to be equi-spaced and the spacing $s$  between the successive binders is treated as a parameter that can attain arbitrary numerical values. In this letter we have presented results only for the special case where $s$ is commensurate with the MT lattice spacing, $\ell$. However, in general, $s$ need not be commensurate with MT lattice spacing \cite{shtylla11}. One can interpret inconmesurate spacing as a more random arrangement of coupler heads. 

If the internal coupler friction increases significantly (measured by $b$), the coupler loses its ability to adapt its position sufficiently fast  with the changes in the position of the MT tip. That is why, in the main text, we refer to the {\it strong friction} regime couplers as being practically static relative to the kMT lattice. Consequently, in this limit, detachment is possible only because of the exit of the kMT tip from the coupler caused by its depolymerization. 


In the {\it in-vitro} experiments \cite{akiyoshi10} the binder heads might show heterogeneous binding energies, or even stochastic fluctuations in the numerical value of $N_{b}$, which are ignored in our simple model. 

\subsection{Microtubule Kinetic rates}
As in the previous work \cite{joglekar02,shtylla11}, we assume that the MT polymerization rate $\alpha(x)$ inside the coupler (if non-zero) is small for all valuees of $x$, except for  $x=L$ where $\alpha =0$ because of the lack of space between the MT tip and the kinetochore wall. On the other hand, we assume that the depolymerization rate $\beta$ of the inserted MT slows down as tension force $F_{load}$ applied to the coupler increases. 
The polymerization and depolymerization rate functions used for the kMT tip kinetic rate $r=\alpha-\beta(F_{\text{load}})$ are given by

\begin{align}
 \alpha(x)&=\frac{\alpha}{1+\exp(-\lambda_{1}(x-\alpha_{1}))},\\
 \beta(F_{\text{load}})&=\beta_{\text{max}} e^{-\lambda F_{\text{load}}} 
 \end{align}

For a given $F_{load}$, the kMT can be in a state of either polymerization ($\alpha>\beta$) or depolymerization ($\beta>\alpha$). For all the calculations shown here, $\alpha$ and $\beta$ were such that the MT was in the state of depolymerisation when $F_{load}=0$. On gradually increasing $F_{load}$, the MT can switch to the state of polymerisation because of tension-induced suppression of $\beta(F)$. 

In this model, we refrain from introducing stochastic transitions between states of polymerization and depolymerization using rescue and catastrophe frequencies. However, such an extension can be easily introduced in the model and we will explore it elsewhere.  Of course, the introduction of stochastic catastrophe/rescue transitions in the kinetic kMT rates will add noise to the mean attachment time calculations. However, as we'll report elsewhere, our key qualitative observations made in this letter are not significantly affected by this noise.

\subsubsection{Load-Independent kMt rates for hybrid couplers.} For this scenario we use
\begin{align}
 \beta(x)&=\beta_{\text{min}}+\frac{\beta_{\text{max}}-\beta_{\text{min}}}{1+\exp(\lambda_{1}(x-\beta_{1}))}
 \end{align}
in agreement with previous work \cite{shtylla11}. We note that in this case, $r=\alpha-\beta$ is adjusted such that the kMT experiences growth up to $x<\beta_{1}=35$ nm, with $\alpha>\beta_{\text{min}}$ and subsequent destabilization for $x>\beta_{1}$ with $\alpha<\beta_{\text{max}}$. The rate $\alpha$ decays to zero if $x>\alpha_{1}=L-\ell$, where $r$ transitions again, with $r=-\beta_{\text{max}}$. In the simulation results shown in Fig. S3 we use $\beta_{\text{min}}=27$ s$^{-1}$, $\beta_{\text{max}}=100$ s$^{-1}$, and $\alpha=80$ s$^{-1}$.

\subsection{Mean Attachment time calculations}

Here we give details of our calculations of the mean lifetime of the kMT-inetochore attachment within the framework of our  model. We take a continuum approach  \cite{redner}.  Given that we start at a position $x$,  
the time required to exit from the boundary $x=0$ is denoted by $T(x)$; it satisfies the delay differential equation
\begin{align}
-1&=\frac{1}{\xi}(-\Psi_{b}'(x) - F)\partial_{x}T(x)+D\partial^2_{x}T(x)+\alpha(x)(T(x+\ell)-T(x))+\beta(x)(T(x-\ell)-T(x))\label{eq6}
\end{align}
As shown in \cite{shtylla11}, in the {\it slippery regime}, the mean first exit time $T(x)$ can be calculated using a reduced ordinary differential equation,
\begin{equation}
V'(x,F,\alpha,\beta)T_{x}(x)+DT_{xx}(x)=-1.\label{exitode}
\end{equation} 
where 
\begin{equation}
\sigma(x)=\sqrt{2D}=\sqrt{\dfrac{2k_{B}T}{\xi}}, 
\end{equation}
\begin{equation}
V'(x,F,\alpha,\beta)=\dfrac{1}{\xi}(-\Psi_{b}'(x)-F)-\ell I_{0}^{2}(f(x)/k_{B}T)\Big(\beta(F)-\alpha(x)\Big) 
\end{equation}
and  $I_{0}(f(x)/k_{B}T)$ is the integral form of the modified Bessel function of the first kind,  which scales the effect of the MT polymerization/depolymerization rates against the activation barrier heights in the potential energy well $\Psi_{b}(x)$. The function $f(x)$, in the Bessel function, arises from a smoothed approximation of the well where we use $\Psi(x)=f(x)(1-\cos(\frac{2\pi x}{\ell})) + h(x)$, with the linear terms $f(x),h(x)$ chosen from a Fourier series fit to the potential energy well, ($f(x)=\frac{b}{2\ell}x +1.5$, $h(x)=-\frac{ax}{\ell}$). This approximation is made in order to simplify the numerical calculations. 

The boundary conditions are $T(0)=0, T'(L)=0$, where we have an absorbing boundary at $x=0$ and a reflecting boundary at $x=L$; the reflecting boundary represents an impenetrable physical barrier erected by  the rigid kinetochore plate. For the mean attachment time, the Eq. S(\ref{exitode}) yields the formal analytical solution 
 \begin{equation}
 T(x)=\frac{1}{D}\int_{0}^x\frac{dy}{\exp(V(y)/D)}\int_{y}^{L}\exp(V(z)/D)dz.
 \end{equation}

To get a closed form expression for the mean first passage time we consider the special case of {\it slippery limit} where the activation barriers against transitions between successive binding sites are sufficiently low ($b \ll k_BT$) such that the potential well is well-approximated by $\Psi(x)=-\hat{a}x$, $\hat{a}=a/\ell$. Further, one can extend the binders to the coupler boundary, such that the well does not become flat inside the coupler (i.e., it remains linear). In this special case the contribution from the Bessel function is  unity ($I_{0}(0)=1$), and the equation for the mean first passage time is simplified to 
 
 \begin{equation}
\Big(\dfrac{\hat{a}-F}{\xi}-\ell \beta_{\text{max}}\exp(-\lambda(F))\Big)T_{x}(x)+DT_{xx}(x)=-1,
\end{equation} 
where we have set $\alpha=0$ without loss of generality. 
Solving this differential equation we get  
 \begin{align}
 T(L)\approx 
 \frac{L^2}{D} \frac{\exp \left(-w\right)-1+w}{w^2}\label{slip_time}
 \end{align}
where 
\begin{align} 
w=\dfrac{L(-\hat{a}+F+\ell \beta_{\text{max}}\xi\exp(-\lambda F))}{k_{B}T} 
\end{align} 
is a non-dimensional quantity. This formula provides a good estimate of the mean attachment times obtained from simulations in the limit $b<<k_{B}T$.

In the {\it strong friction} regime, diffusion of the coupler along the kMT is insignificant and binding is very strong ($a,b>>k_{B}T$), so that movement of the kMT inside the coupler is only accomplished with the Poisson counting processes. One can alternatively think of this regime as a velcro or sticky regime, where the coupler is so strongly engaged with kMT binding sites that relocation relative to the kMT is not possible even under load, due to the high binding energy  (coupler binding will eventually break under significant load, however for this limit coupler breaking would require load beyond the ranges tested {\it in vitro}).

In this case,  
 \begin{equation}
 dx(t)  \simeq \ell dN_{r}(t)
 \end{equation}
 In the special case of a MT with $\alpha=0$, the overlap inside the coupler depends only on the depolymerization steps counted by the process $dN_{\beta(F)}$.
 For this process, the time interval between the occurrence of the consecutive depolymerization events are independent identically distributed random variables with mean $1/\beta(F)$. 
Therefore, the mean exit time for a fully engaged coupler with length $L=N_{tot}\ell$ in the strongly bound regime is
 \begin{equation}
  T(L)\approx\dfrac{N_{tot}}{\beta_{\text{max}} e^{-\lambda F_{\text{load}}}}=\dfrac{L}{\ell\beta_{\text{max}} e^{-\lambda F_{\text{load}}}}.
  \end{equation}
  
\subsection{dad1-1 deletion in silico and force-velocity calculations}

\begin{figure}[htbp]
\begin{center}
\includegraphics[width=4in,height=2.33in]{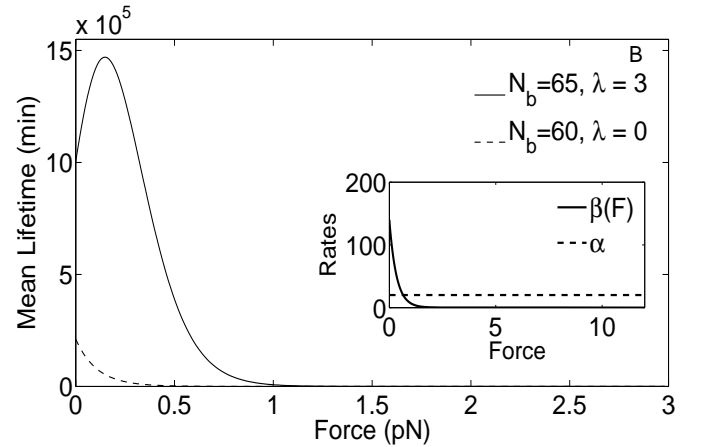}
\caption{\small  Mean attachment times in the slippery regime. We show calculations with $N_{b}=65$ (solid line) and $N_{b}=60$ (dashed line), using eq.~(S\ref{slip_time}). We also remove force dependent depolymerization in one case ($\lambda=0$), which completely removes range of forces for which attachment times increase. These results are used to reproduce $Dad1$-1 deletion studies from \cite{akiyoshi10}. {\it Figure inset.} Plot of the kMT kinetic rates as a function of load force. Common parameters for all calculations shown are $\lambda=3$ pN$^{-1}$, $\beta_{max}=140$, $\alpha=20$, $a=0.4$ k$_{B}$T, $k=0.001a$.}
\label{fig6}
\end{center}
\end{figure}

Force dependent depolymerization is key in order to observe the non-monotonic variation of the mean lifetime of the coupler with the load force. We test our model by reproducing the results of the {\it in-vitro} experiments reported in \cite{akiyoshi10}, where $dad1-1$ components were deleted from the coupler. To mimic these deletions, we remove $5$ binders and we set $\lambda=0$ in order to remove lateral coupler cohesion provided by DAM ring components. In this scenario, the mean attachment times decrease monotonically with increasing load force (see Fig S\ref{fig6}) in full agreement with the corresponding observation of Akiyoshi et al. \cite{akiyoshi10}.  

\subsection{Velocity Reversals}

A suitably chosen value of the parameter $N_{b}$ or $a$ can ensure that the mean attachment times exceed $1$ min for the range of $F_{load}$ for which velocities are measured by simulations of eq. (1). In the inset of Fig.~ 3, we show coupler speeds for such a stably attached coupler in the {\it slippery} regime. The typical values of these speeds are comparable to the data of  ref.\cite{akiyoshi10}. 

 
In the plot of the speed versus force, shown in  Fig.~3inset, we see that the coupler experiences first a gradual slow-down and, beyond the velocity transition point,  eventual speeding up as the tension load is increased further. The transition point corresponds to $F_{c}=1/\lambda=1/3$ pN, which is consistent with the form of the $\beta(F_{\text{load}})$ function. We highlight that the value of $F_{c}$ can be increased only in the cases where there is sufficient $a,N_{b}$ to support attachment; in order for the nonlinearities to persist it is important that the depolymerization rate is slowed down within the range of forces that can be supported by the coupler.

\subsection{Numerical Simulations}

The parameters in the ``intermediate regime'' satisfy neither the conditions for {\it slippery} motion  nor those for {\it strong friction}. In this parameter regime the mean lifetime of the kMT-kinetochore 
attachments were computed by computer simulations of eq. (1) of the main text. The averaging was carried out over 500 trials for each fixed value of the force. Time step size had to be sufficiently small to ensure numerical stability and convergence. Exit time searches were extended up to 60 min, which is well within the range of times explored experimentally in \cite{akiyoshi10}. 
In the stably attached cases, with high $a,N_{b}$ (or large attachment energy), numerical trials for exit times failed to yield an estimate, because the coupler remained engaged for times exceeding the allotted 60 min. Such an artificial truncation of the lifetime distribution at 60 min gives rise to artificially lower estimates for the mean and standard deviation of attachment times at the exit time peak points in the intermediate regime; this is due to the smaller number of points used to gather exit time statistics.

\subsection{Active Coupler Interface}
Here we describe our approach for incorporating force-generation by active components at the coupler interface into the dynamical equations. We will refer to these elements as generic molecular motors, however, we note that in the case of kMT minus end motors ($n_{-}$), then the force of the motors is equivalent to the force exerted by a ring that is being pushed by the curling plus end tips of kMTs.  We note here that minus end motors will increase the kMT-coupler overlap, because these components walk toward the MT minus end, while the plus end of the kMT is the side that gains attachment with the coupler. On the other hand, for plus end motors, their action will decrease kMT-coupler overlap because the motors push toward the plus-end tip of the inserted kMT.

For our purposes, we find it sufficient to adopt a standard prescription for capturing the force-generation by the molecular motors that does not explicitly describe the detailed stochastic mechano-chemistry of the individual motors. 

\subsubsection{Derivation of the model equations for the coupler composed of binders and motors.}
We start with force balance equation which does not include random fluctuations for the coupler overlap velocity
\begin{equation}
 \frac{dx(t)}{dt}-V_{MT}=\frac{1}{\xi}\sum F=\frac{1}{\xi}\Big(-\Psi_{b}'(x)-F_{\text{load}}+ F_{\text{A}}(x)\Big)\label{determ_velocity},
\end{equation}
where $\xi$ is the coupler effective drag coefficient and $V_{MT}$ is the velocity of the kMT tip with respect to a space-fixed frame of reference. The active force term
\begin{equation}
F_{\text{A}}(x)=d_{m}(x)(n_{-}f_{-}-n_{+}f_{+})
\end{equation}
with the motor density function
\begin{align}
d_{m}(x)&=(x-N_{b}s)(H(x-N_{b}s)-H(x-N_{b}s-L_{m}))\\
&+ L_{m} H(x-N_{b}s-L_{m}),
\end{align}
where $H(x)$ is the standard Heaviside step function and $L_{m}=8$ nm corresponds to the total horizontal length of the coupler that can be populated by active components (in three-dimensions this corresponds to one layer of motors working around a kMT with 12 protofilament tracks, with one motor per track).

As noted in the main text, linear force-velocity relations permit us to explicitly calculate the active component velocity dependence for the total active components using  $f_{-}$ and $f_{+}$. Note that minus here denotes minus end directed motors/or protofilament curling that push to increase overlap and plus denotes plus end directed motors that work against coupler/microtubule overlap. For each case, we have
\begin{align}
f_{\pm}=F_{max}^{\pm}\left( 1 - \frac{v_{\pm}}{V^{\pm}_{max}}\right),
\label{eq-fplusminus}
\end{align}
where $F_{max}^{\pm}$ and $V_{max}^{\pm}$ are the stall force and maximal velocity for the plus-end directed and minus-end directed motors, respectively, whereas $v_{\pm}$ are the corresponding instantaneous velocities. 
Next we express $v_{\pm}$ in terms of $dx/dt$.  Ignoring all the binder fibers,  $x=x_{tip}-x_{motor}$ and hence
\begin{align}
\frac{dx}{dt}&=\frac{dx_{tip}}{dt}-\frac{dx_{motor}}{dt}\\
&=V_{MT}+v_{-}
\label{eq-vminus}
\end{align}
Note that, in the absence of plus-end-directed motors and ignoring boundary conditions, the overlap can attain a stationary value only if $V_{MT}<0$; in this stationary state the depolymerization of the tip would be balanced by the translocation driven by the minus-end-directed motors. Similarly, if only plus-end-directed motors are present,
\begin{align}
\frac{dx}{dt}&=V_{MT}-v_{+}.
\label{eq-vplus}
\end{align}
Substituting eqs.~(\ref{eq-vminus}) and (\ref{eq-vplus}) into eq.~(\ref{determ_velocity}) we get
\begin{align}
\frac{dx}{dt}=&\nonumber \\
\frac{1}{\xi}&\Big[ -\Psi'(x)-F_{\text{load}} + d(x)(n_{-}F_{max}^{-}\left( 1- \frac{dx/dt-V_{MT}}{V^{-}_{max}}\right)\nonumber\\
&-n_{+}F_{max}^{+}\left( 1- \frac{V_{MT}-dx/dt}{V^{+}_{max}}\right)\Big]+V_{MT}.
\end{align}

Regrouping the velocity terms we obtain the following equation for coupler overlap,


\begin{align}
\frac{dx}{dt}&=\frac{1}{\xi(x)}\left[ -\Psi'(x)-F_{\text{load}} + d(x)(n_{-}F_{max}^{-}-n_{+}F_{max}^{+})\right]\nonumber\\
&+V_{MT}.
\label{deteqn}
\end{align}
where $\xi(x)=\xi+\mu^{-}(x)+\mu^{+}(x)$ and
\begin{equation}
\mu^{\pm} = d(x)\biggl(\frac{n_{\pm} F^{\pm}_{max}}{\xi V^{\pm}_{max}}\biggr). 
\end{equation} 

\begin{figure}[hbtp]
\begin{center}
\includegraphics[width=3.5in,height=2.1in]{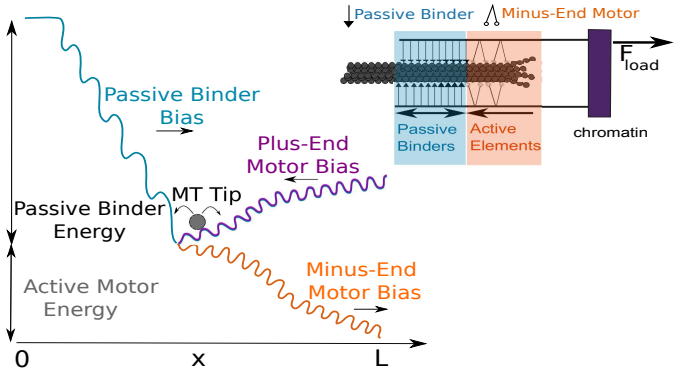}
\end{center}
\caption{\small 
Potential binding energy modification upon active interface addition.
}\label{fig4}
\end{figure}

We note that the potential well diagram in Fig. S\ref{fig4} demonstrates the effects of incorporating active components in the couplers. For minus-end motors the well is tilted in such a way so that the minimum energy state is located at $x=L$, for which the coupler is fully overlapped with the kMT. Notice that the shape of the well indicates an increased internal friction arising not just due to the passive binder heads (local wells) but also due to the concavity of the modified well due to the minus end motors. On the other hand, for plus end motors cause the minimum energy state to correspond to $x=N_{b}s$, which places the passive binders at the kMT tip. This effect causes the coupler to operate under a biased diffusion principle, especially when external loads are tension loads ($F_{\text{load}}>0$).

The dynamics of the coupler described by eq.~(\ref{deteqn}) is fully deterministic. However, in reality, the kinetics of the kMT-kinetochore coupler are stochastic. Therefore, we now write down a stochastic differential equation (SDE) that would, upon averaging, correspond to the deterministic equations written above. Suppose over a small time interval $\delta t$ the number of subunits (an $\alpha-\beta$ tubulin dimer) added and removed from the tip of the kMT by polymerization and depolymerization are $dN_{r}$, an independent homogenous Poisson process. 

Moreover, we capture the effects of random Brownian forces through the noise $W(t)$ which is assumed to be a Gaussian stochastic process. We distinguish this Gaussian process from the one used earlier in the absence of active force generators. Since molecular motors are fueled by chemical reactions (e.g., ATP hydrolysis), this random force includes the effects of fluctuations both in the chemical reactions and mechanical stepping involved in each cycle of the individual motors. 

Thus, the equation for the coupler overlap reads
\begin{align}
\frac{dx}{dt}&=\frac{1}{\xi(x)}\left[ -\Psi'(x)-F_{\text{load}} + d(x)(n_{-}F_{max}^{-}-n_{+}F_{max}^{+})\right]\nonumber\\
&+\ell dN_{r}(t)+dW(t).
\end{align}

\subsection{Force velocity relations for a hybrid coupler}

Here we show force-velocity calculations for a coupler motor that is engaged with varying densities of plus-end directed or opposing motors and the inserted kMT kinetic rates do not have load dependence.

\begin{figure}[htbp]
\begin{center}
\includegraphics[width=3.in,height=2.2in]{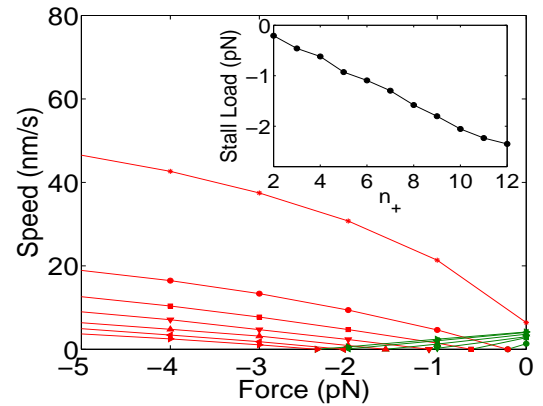}
\caption{Average speed versus force plot for a stably bound coupler with varying densities of plus-end directed motors. {\it Inset}. Stall force calculations for each motor density inside the coupler. The parameters are $N_{b}=52$, $\lambda=3$, $a=$ k$_{B}$T, $k=0.01a$, $\beta_{\text{min}}=27$ s$^{-1}$, $\beta_{\text{max}}=100$ s$^{-1}$, and $\alpha=80$ s$^{-1}$. }
\label{figspeed}
\end{center}
\end{figure}

In order to challenge active components that work to decrease the overlap (such as CENP-E kinesin), in Fig.~S\ref{figspeed} we tested the model with varying $n_{+}$ and $F_{\text{load}}<0$. In these cases, $F_{\text{load}}<0$ is not a tension load, so we omit kMT $\beta$ dependence on $F_{\text{load}}$. As shown in Fig. S\ref{figspeed},  when the active motor component pulls to oppose coupler overlap against compressing load, the internal friction is directly increased by the action of the motor, resulting in an overall loss of coupler tracking ability (however attachment is maintained, the motors become stalled), as shown by decreasing velocities in Fig. S\ref{figspeed}. 


\section{Error correction at kinetochores and catch-bond-type mechanisms}
The non-monotonic attachment times which arise for various parameter ranges for this coupler model indicate that there are force ranges for which longer attachment times are favored. We interpret the regions with peak attachment times as stable-attachment force regions. Thus the force ranges that give long $T(L)$ can be thought of as optimal force ranges for which kinetochores support attachment. This force-selective mechanism is particularly important in the context of error correction of kMT attachments at kinetochores; many erroneous kMT attachments may not provide sufficient tension force on a kinetochore due to the geometry of the connection. A purely force-mediated error-correction mechanism at kinetochores has important implications because it expands the role of kinetochore couplers beyond generation of movement, to also checking the quality of the kMT attachment.

There are various mechanisms for catch-bonds that explain how such a bond can become stronger under force (for review see \cite{thomas08}). In the general context of MT-kinetochore attachments, 
references to 'catch-bonds' are made to account for the observed increase in the attachment  life time when subjected to tensile force.  We have offered a conceptual scenario where the stronger is the tension the smaller is the curvature of the splaying kMT tips and, as a consequence, the slower is the depolymerization of the kMT and, hence, the longer is the life time. 

To implement this conceptual model quantitatively, we have assumed that the depolymerization rate is proportional to the free depolymerization rate (or attempt frequency) and exponentially related to the height of the energy barrier along the unbinding pathway. We have decoupled the coupler binding energy by keeping the energy function $\Psi(x)$ and modified the depolymerization rate by a rescaled load force term, $F_{\text{load}}\lambda$. The two can be combined if we use a dynamic potential energy landscape which takes into account protofilament curling energies ($\Psi(x,t)$), however, we claim that the end effect is similar to the simple model we present here.

\clearpage
\section{Parameter Values}

\begin{table}[htdp]
\begin{center}
\begin{tabular}{ccc}

Parameter Description & Symbol & Values Tested \\
\hline
\hline \cr
MT binding site spacing & $\ell$ & 8/13 nm \cite{hill85b}\cr\\
Maximal coupler length &$L$& 50 nm \cite{joglekar06}\cr \\
Maximal number of coupler binders &$N_{b}$& 15-65 \cr \\
Maximal Depolymerization Rate & $\beta_{max}$ & $100- 350$ $s^{-1}$ \cite{hill85b}\cr \\
Maximal Polymerization Rate & $\alpha$ & $20-50$ $s^{-1}$\cite{joglekar02}\cr \\
MT lattice/binder binding energy & $a$ & $0.4k_{B}T-3 k_{B}T$\cr\\ 
Binder activation barrier   & $b$ & $0.001a-0.4a$\cr\\
Critical Depolymerization Force   & $F_{c}=1/\lambda$ & $0.3-5$ pN\cr\\
Polymerization Decay Position   & $\alpha_{1}$ & $L-\ell$ nm \cite{shtylla11}\cr\\
kMT-rate transition steepness   & $\lambda_{1}$ & $100$ nm$^{-1}$ \cite{shtylla11}\cr\\
\hline
\end{tabular}
\end{center}
\label{default}
\end{table}%

\end{document}